\documentclass[a4paper]{jpconf}
\usepackage{graphicx,iopams}
\begin{document}

\begin{flushright}
{\small MITP/15-006\\ IFT-UAM/CSIC-15-008
}
\end{flushright}

\title{Discrete Abelian gauge symmetries and axions}  

\author{\underline{Gabriele Honecker}${}^1$ and Wieland Staessens${}^2$}

\address{${}^1$ Cluster of Excellence PRISMA \& Institute of Physics (WA THEP), Johannes Gutenberg University, 55099 Mainz, Germany}
\address{${}^2$ Instituto de F\'isica Te\'orica UAM-CSIC, Cantoblanco, 28049 Madrid, Spain}

\ead{${}^1$ Gabriele.Honecker@uni-mainz.de; 
${}^2$ wieland.staessens@csic.es}

\begin{abstract} 
We combine two popular extensions of beyond the Standard Model physics within the framework of intersecting D6-brane models: discrete $\mathbb{Z}_n$ symmetries and Peccei-Quinn axions. The underlying natural connection between both extensions is formed by the presence of massive $U(1)$ gauge symmetries in D-brane model building. Global intersecting D6-brane models on toroidal orbifolds of the type $T^6/\mathbb{Z}_{2N}$ and $T^6/\mathbb{Z}_{2} \times  \mathbb{Z}_{2M}$ with discrete torsion offer excellent playgrounds for realizing these extensions. A generation-dependent $\mathbb{Z}_2$ symmetry is identified in a global Pati-Salam model, while global left-right symmetric models give rise to supersymmetric realizations of the DFSZ axion model. In one class of the latter models, the axion as well as Standard Model particles carry a non-trivial $\mathbb{Z}_3$ charge. 
\end{abstract}

\section{Introduction}\label{S:Intro}

Extensions of the Standard Model are characterised by the inclusion of new particles to resolve open questions arising in particle physics and cosmology. 
These new particles are subject to novel (gauge) symmetries which constrain the interactions between the observed particles and the yet to be discovered ones. Obviously the most simple new gauge symmetries to appear in beyond the Standard Model scenarios are local Abelian symmetries.
As it turns out, Abelian gauge symmetries are ubiquitous in four-dimensional string theory vacua, and within Type II  string theory compactifications they appear in three species:
\begin{itemize}
\item
closed string vectors associated to isometries of the compact six-dimensional space,
\item
massless open string vectors with matter charged under them,
\item
massive open string vectors with masses of the order of $M_{\rm string}$.
\end{itemize}
It has only been realized recently that the massive vectors do not decouple completely from the low-energy effective field theory, but that discrete subgroups can survive~\cite{BerasaluceGonzalez:2011wy,Ibanez:2012wg,Honecker:2013hda,Honecker:2013kda,Berasaluce-Gonzalez:2013sna,Marchesano:2013ega} and act as ultimate selection rules for  non-perturbative instantonic $m$-point couplings. 
This observation complements the discussion of  discrete gauge symmetries in field theory invoked to guarantee e.g. the stability of the proton~\cite{Ibanez:1991hv,Ibanez:1991pr,Dreiner:2005rd,Lee:2011dya}.

On the other hand, the massive Abelian vectors lead to global $U(1)$ symmetries in perturbation theory, which might be broken explicitly by vacuum expectation values of light charged open string states. This is exactly the stringy generalization of a field theoretical Peccei-Quinn symmetry. The relevant charged pseudo-Nambu Goldstone bosons or {\it axions} stem from the open string sector, whereas the CP-odd pseudoscalars or {\it axions} that are absorbed by massive vectors to form their longitudinal modes stem from the closed string sector.\footnote{In particle physics phenomenology, pseudoscalars are often called {\it axion-like particles (ALPs)}, while the terminology {\it axion} is reserved for the {\it QCD axion} invoked to solve the strong CP problem.}
In contrast to purely field theoretical models, the origin of axions from the open string sector severely constrains the possible realization of a Peccei-Quinn symmetry~\cite{Honecker:2013mya} as discussed in detail in section~\ref{S:OpenAxions}. Hence, the study of $U(1)$ symmetries within Type II string theory offers an immediate possibility to discuss another popular extension of the Standard Model: Peccei-Quinn axions.    

The discussion in this article concentrates on Type IIA orientifolds, but by mirror symmetry analogous results hold in Type IIB orientifolds, see e.g.~\cite{Garcia-Etxebarria:2014qua,Mayrhofer:2014haa,Mayrhofer:2014laa,Karozas:2014aha} for discrete symmetries in the strong coupling version, F-theory.

\section{Closed string axions and the Green Schwarz mechanism}\label{S:Axions+GS}

The closed string sector of Type IIA/$\Omega{\cal R}$ orientifold theories with an anti-holomorphic involution ${\cal R}$ along three complex compact dimensions accompanying the worldsheet parity $\Omega$ contains several types of axions as complexifications of geometric K\"ahler and complex structure moduli as well as the dilaton, see the reviews~\cite{Blumenhagen:2006ci,Ibanez:2012zz}. The complete list of massless bosonic closed string states and their allocation to $\mathcal{N}=1$ supersymmetric multiplets is summarized in table~\ref{Tab:closed-IIAspectra}.
\begin{table}[h!]
\caption{\label{Tab:closed-IIAspectra} Multiplicities of closed string multiplets in four-dimensional Type IIA/$\Omega{\cal R}$ models.}
  \begin{center}
\begin{tabular}{ccc}\br
${\cal N}=1$ multiplet & \# & bosons \\\mr
gravity & 1 & $(G_{\mu\nu})$\\
linear (dilaton-axion) & 1 & $(\phi,\xi_0)$ \\
chiral (complex structures) & $h_{21}$ & $(c_k,\xi_k)$\\
chiral (K\"ahler moduli) & $h_{11}^-$ & $(v_i,b_i)$\\
vector & $h_{11}^+$ & $(A_i)$
\\\br
\end{tabular}
\end{center}
\end{table}

The axionic partners $\xi_{i, i \in \{0 \ldots h_{21}\}} \equiv \ell_s^{-3} \int_{\Pi_i^{\rm even}} C_3^{\rm RR}$ of the dilaton and complex structure moduli  and the corresponding four-dimensional Hodge dual two-forms ${\cal B}_2^{(i)} \equiv \ell_s^{-5} \int_{\Pi_i^{\rm odd}} C_5^{\rm RR}$ 
couple to gauge fields via the Chern-Simons action of the associated D6-branes,
\begin{equation}\label{Eq:GS-couplings}
\mathcal{S}_{\rm CS}  \supset \int_{\mathbb{R}^{1,3}} \sum_{i=0}^{h_{21}} \left( 
B^i_a \; {\cal B}^{(i)}_2 \wedge {\rm tr}F_a + A^i_b \;  \xi_{i} \; {\rm tr} F_b \wedge F_b  
\right),
\end{equation}
with coefficients $B^i_a, A^i_b \in \mathbb{Q}$ such that all mixed and purely Abelian field theoretical anomalies cancel.
A linear combination of Abelian gauge factors $U(1)_X=\sum_a q_a U(1)_a$ with coefficients $q_a \in \mathbb{Q}$
 remains massless if all its couplings on the left hand side of equation~(\ref{Eq:GS-couplings}) vanish, i.e. $\sum_a N_a q_a  B^{(i)}_a=0$ for all $ i \in \{0, \ldots, h_{21}\}$ and the factor $N_a$ for $U(1)_a \subset U(N_a)$.
This is exactly the set-up required for a massless hypercharge or gauged $(B-L)$ symmetry.

The remaining Abelian gauge symmetries with $\sum_a N_a q_a  B^{(i)}_a \neq 0$ for some $i$ acquire masses at the string scale $M_{\rm string}$, but remain as global symmetries in perturbation theory, which are broken by non-perturbative effects such as instantons in the full string compactification, see e.g.~\cite{Blumenhagen:2009qh}, as required by general arguments on the non-existence of global continuous symmetries in the presence of gravity, see e.g.~\cite{Kallosh:1995hi}.

\section{Discrete subgroups of massive U(1)s}\label{S:Discrete}

In Type II string perturbation theory, axionic partners of the dilaton and compactification moduli only appear with derivatives in the Lagrangian, $ \sum_{i=0}^{h_{21}} (\partial_{\mu} \xi_i)(\partial^{\mu} \xi_i)\subset \mathcal{L} $, due to the special structure of the closed string K\"ahler potential, $\mathcal{K}_{\rm closed} = - \ln (\phi) - \sum_{k=1}^{h_{21}} \ln c_k - \sum_{i=1}^{h_{11}} \ln v_i$, which only contains the dilaton and geometric moduli. The global continuous shift symmetry is in the Type IIA formulation generically broken to $\xi_i \to \xi_i +1$ by D2-brane instanton effects which scale as $e^{-{\cal S}_{\rm inst}}$
 with 
 \begin{equation}\label{Eq:S-instanton}
 {\cal S}_{\rm inst} \supset 2\pi {\rm i} \sum_i A^i_{\rm D2} \xi_i 
\end{equation} 
 with $A^i_{\rm D2} \in \mathbb{Q}$ the same coefficients as for a stack of D6-branes $b$ on the right hand side of equation~(\ref{Eq:GS-couplings}) wrapping the identical compact three-cycle $\Pi_b$ as the D2-brane instanton.  

By the Hodge duality relations $d{\cal B}_2^{(i)} = m_i \ast_4 d\xi_i$ with yet undetermined constants $m_i \in \mathbb{Z}$, the left hand coupling in equation~(\ref{Eq:GS-couplings})
leads to the following shift in the axion upon a gauge transformation of some massive $U(1)=\sum_a k_a U(1)_a$ vector boson $\mathcal{A}_{\mu}$,
\begin{equation}\label{Eq:Discrete_GaugeTrafo}
 \xi_i \rightarrow \xi_i + \Bigl(\sum_a N_a k_a B_a^i m_i \Bigr) \lambda,
 \hspace{0.3in}
\mathcal{A}_\mu \rightarrow \mathcal{A}_\mu + \partial_\mu \lambda  .
\end{equation} 
Instanton contributions to the effective action are now invariant under a discrete $\mathbb{Z}_n$ subgroup of gauge transformations if 
\begin{equation}\label{Eq:Zn-condition}
\sum_a N_a k_a B_a^i m_i  \stackrel{!}{=} 0 \; {\rm mod } \; n \qquad {\rm for \; all} \; i =0 \ldots h_{21},
\end{equation} 
as can be read off from the instanton action in equation~(\ref{Eq:S-instanton}).

It remains to determine the constants $m_i$, which depend on the choice of the compact background as follows.
Any three-cycle $\Pi_a $ and its image $\Pi_a^{\prime}$ under the anti-holomorphic involution ${\cal R}$   can be expanded in terms of even and odd components $\Pi^{\rm even/odd}_i$ with coefficients $A^i_a, B^i_a$,
\begin{equation}\label{Eq:even+odd-cycles-expansion}
\Pi_a = \sum_{i=0}^{h_{21}} \Bigl( A^i_a \, \Pi^{\rm even} + B^i_a\, \Pi^{\rm odd} \Bigr),
\qquad
\Pi_a^{\prime}= \sum_{i=0}^{h_{21}} \Bigl( A^i_a \, \Pi^{\rm even} - B^i_a\, \Pi^{\rm odd} \Bigr)
.
\end{equation} 
In the most simple case discussed in~\cite{BerasaluceGonzalez:2011wy}, where the unimodular lattice of three-cycles is aligned with the $\mathcal{R}$-invariant direction, one has $\Pi^{\rm even}_i \circ \Pi^{\rm odd}_j = \pm \delta_{ij}$ and $m_i = \pm 1$.

However, for all phenomenologically appealing D6-brane vacua found on $T^6$ and $T^6/(\mathbb{Z}_2 \times \mathbb{Z}_2)$ without discrete torsion $(\eta=+1)$~\cite{Ibanez:2001nd,Cvetic:2001tj,Cvetic:2001nr}  or $T^6/\mathbb{Z}_{6}^{(\prime)}$~\cite{Honecker:2004kb,Honecker:2004np,Bailin:2006zf,Gmeiner:2007we,Bailin:2007va,Gmeiner:2007zz,Bailin:2008xx,Gmeiner:2008xq,Gmeiner:2009fb} or $T^6/(\mathbb{Z}_2 \times \mathbb{Z}_6^{(\prime)})$ with discrete torsion $(\eta=-1$)~\cite{Forste:2010gw,Honecker:2012qr,Honecker:2013kda,Ecker:2014hma,Ecker:2015jan} and generally for arbitrary Calabi-Yau manifolds, the lattice spanned by $\{\Pi^{\rm even}_i, \Pi^{\rm odd}_j\}$ with integer coefficients only forms a sublattice $\Lambda^{\rm even}_3 \oplus \Lambda^{\rm odd}_3  \subsetneq \Lambda_3$ of finite index of  the full lattice of three-cycles. The intersection numbers in general take the form 
\begin{equation}\label{Eq:def-m-integer}
\Pi^{\rm even}_i \circ \Pi^{\rm odd}_j = m_i \,  \delta_{ij}
\quad
{\rm with}
\quad m_i \in \mathbb{Z},
\end{equation} 
and the coefficients $A^i_a, B^i_a$ in the expansion~(\ref{Eq:even+odd-cycles-expansion}) correspondingly take rational values in $\frac{1}{m_i} \mathbb{Z}$, which need to be worked out on a case-by-case basis as exemplified in section~\ref{Ss:Orbifolds}.

In order to unambiguously define the conditions on the existence of a discrete $\mathbb{Z}_n$ symmetry, it is useful to rewrite the conditions in equation~(\ref{Eq:Zn-condition})
in terms of intersection numbers with the $(h_{21}+1)$ dimensional basis of $\mathcal{R}$-even three-cycles $\Pi^{\rm even}_i$,
\begin{equation}\label{Eq:discrete-as-intersection}
\Pi^{\rm even}_i \circ \sum_a N_a k_a \Pi_a  =0 \; {\rm mod} \; n  \; \forall i
\quad
{\rm with}
\quad 
k_a \in \mathbb{Z},
\quad 
0 \leqslant k_a <n,
\quad
{\rm gcd}(k_a, k_b,\ldots,n)=1.
\end{equation} 
The range of $k_a$ ensures the uniqueness of weight assignments which are contained in the massive Abelian vector, while the condition on the trivial greatest common divisor ensures that the order $n$ of the discrete symmetry is minimal.
As a cross-check, one can verify that in models where $USp(2)_i$ gauge groups are found for probe D6-branes on any three-cycle $\Pi^{\rm even}_i$, the K-theory constraint corresponds to the $\mathbb{Z}_2$ symmetry with $(k_a,k_b,\ldots)=(1,1,\ldots)$. 

It should be noted that solutions to~(\ref{Eq:discrete-as-intersection}) can appear redundant in four-dimensional field theory. This happens in particular for any $(\underline{k_a,k_b,k_c\ldots})=(\underline{1,0,0\ldots})$ with underlining denoting all possible permutations of entries, whose solution is $\mathbb{Z}_N \subset U(N)$. Even though this $\mathbb{Z}_N$ symmetry does not correspond to the center of $SU(N)$, the charge assignment of any open string in the fundamental $({\bf N})_1$ or antisymmetric $({\bf Anti})_2$ representation of $SU(N) \times U(1) \simeq U(N)$ or the respective conjugates $(\overline{\bf N})_{-1 = N-1 \, {\rm mod } \, N}$ and $(\overline{\bf Anti})_{-2 = N-2 \, {\rm mod } \, N}$ is so constraining that the $\mathbb{Z}_N$ selection rules on couplings do not provide any constraints beyond those imposed by the non-Abelian $SU(N)$ representations.
Moreover, if there exists some massless Abelian gauge symmetry such as the  hypercharge or a gauged $(B-L)$ symmetry
 in the model, it can be used to `rotate away' some $\mathbb{Z}_n$ charges, as we will see in some examples in section~\ref{Ss:models+axions}.

\subsection{Toroidal Orbifolds}{\label{Ss:Orbifolds}

For $T^6/\mathbb{Z}_{2N}$ and $T^6/\mathbb{Z}_2 \times \mathbb{Z}_{2M}$ with discrete torsion $(\eta=-1)$, fractional three-cycles are given by linear combinations from the bulk and $\mathbb{Z}_2$ twisted sectors,
\begin{equation}\label{Eq:def-fractional-cycle}
\Pi_a^{\rm frac} = \frac{1}{2} \left( \Pi_a^{\rm bulk} + \Pi_a^{\mathbb{Z}_2} \right)
\qquad
{\rm or}
\qquad
\Pi_a^{\rm frac} = \frac{1}{4} \left( \Pi_a^{\rm bulk} + \sum_{k=1}^3 \Pi_a^{\mathbb{Z}_2^{(k)}} \right),
\end{equation} 
with each $\Pi_a^{\mathbb{Z}_2^{(k)}}$ consisting of $2 \times 2$ contributions of $\mathbb{Z}_2^{(k)}$ fixed points on $T^4_{(k)} \equiv T^2_{(i)} \times T^2_{(j)}$ times a one-cycle 
on the $\mathbb{Z}_2^{(k)}$-invariant $T^2_{(k)}$.
The Hodge numbers for some phenomenologically favourable backgrounds, which include a $\mathbb{Z}_3$ symmetry, are summarized in table~\ref{Tab:Hodge-numbers},  
\begin{table}[ht!]
\caption{\label{Tab:Hodge-numbers}Hodge numbers $\left(\!\!\begin{array}{c} h_{11} \\h_{21}\end{array}\!\!\right)$ for some phenomenologically appealing toroidal orbifolds. }
  \begin{center}
\begin{tabular}{ccccccccccc}\br
$T^6/$ & \multicolumn{8}{c}{$SU(2)^2 \times SU(3)^2$}
 \\\mr
 sector & bulk
 & $(0,\frac{1}{6},\frac{-1}{6})$ & $(0,\frac{1}{3},\frac{-1}{3})$ & $(0,\frac{1}{2},\frac{-1}{2})$ & $(\frac{1}{2},\frac{-1}{2},0)$ & $(\frac{-1}{2},\frac{1}{3},\frac{1}{6})$ & $(\frac{-1}{2},\frac{1}{6},\frac{1}{3})$ &  $ (\frac{1}{2},0,\frac{-1}{2})$ &  
\\\mr
$\mathbb{Z}_6'$ & $\left(\begin{array}{c} 3 \\ \framebox{1} \end{array}\right)$ 
& & $\left(\begin{array}{c} 12 \\ 6 \end{array}\right)$ & & 
&  \multicolumn{2}{c}{$\left(\begin{array}{c} 12 \\ 0 \end{array}\right)$} 
& $\left(\begin{array}{c} 8 \\ \framebox{4} \end{array}\right)$
\\\mr
 $\begin{array}{c}  \mathbb{Z}_2 \times \mathbb{Z}_6 \\ \eta=-1 \end{array}$& \hspace{-6mm}
 $\left(\begin{array}{c} 3 \\ \framebox{1} \end{array}\right)$ 
 & $\left(\begin{array}{c} 0 \\ 2 \end{array}\right)$
 & $\left(\begin{array}{c} 8 \\ 2 \end{array}\right)$
 & $\left(\begin{array}{c} 0 \\ \framebox{6} \end{array}\right)$
 & $\left(\begin{array}{c} 0 \\ \framebox{4} \end{array}\right)$
 & $\left(\begin{array}{c} 4 \\ 0 \end{array}\right)$
 & $\left(\begin{array}{c} 4 \\ 0 \end{array}\right)$
 & $\left(\begin{array}{c} 0 \\ \framebox{4} \end{array}\right)$
 \\\br
$T^6/$ &  \multicolumn{8}{c}{ $SU(3)^3$}
 \\\mr
 sector & bulk
&  $(\frac{-1}{3},\frac{1}{6},\frac{1}{6})$ & $(\frac{-2}{3},\frac{1}{3},\frac{1}{3})$ & $(0,\frac{1}{2},\frac{-1}{2})$ & $(\frac{1}{2},\frac{-1}{2},0)$ & $(\frac{1}{6},\frac{-1}{3},\frac{1}{6})$ & $(\frac{1}{6},\frac{1}{6},\frac{-1}{3})$  &  $(\frac{1}{2},0,\frac{-1}{2})$ &  
\\\mr
$\mathbb{Z}_6$ &  $\left(\begin{array}{c} 5 \\  \framebox{0} \end{array}\right)$
& $\left(\begin{array}{c} 3 \\  0 \end{array}\right)$
& $\left(\begin{array}{c} 15 \\  0 \end{array}\right)$
& $\left(\begin{array}{c} 6 \\  \framebox{5} \end{array}\right)$
\\\mr
$\begin{array}{c}  \mathbb{Z}_2 \times \mathbb{Z}_6' \\ \eta =-1  \end{array}$ \hspace{-6mm}
&$\left(\begin{array}{c} 3 \\  \framebox{0} \end{array}\right)$
& $\left(\begin{array}{c} 1 \\  0 \end{array}\right)$
&  $\left(\begin{array}{c} 9 \\ 0 \end{array}\right)$
&  $\left(\begin{array}{c} 0 \\ \framebox{5} \end{array}\right)$
&  $\left(\begin{array}{c} 0 \\ \framebox{5} \end{array}\right)$
&  $\left(\begin{array}{c} 1 \\ 0 \end{array}\right)$
&  $\left(\begin{array}{c} 1 \\ 0 \end{array}\right)$
&  $\left(\begin{array}{c} 0 \\ \framebox{5}  \end{array}\right)$
\\\br
\end{tabular}
\end{center}
\end{table}
where the relevant contributions to $h_{21}$ from the bulk and $\mathbb{Z}_2$ twisted sectors are highlighted in boxes. For $\mathbb{Z}_6$ and $\mathbb{Z}_2 \times \mathbb{Z}_6'$, the full unimodular lattice of three-cycles is available for model building, whereas for $\mathbb{Z}_6'$ and $\mathbb{Z}_2 \times \mathbb{Z}_6$ only the sublattice spanned by the bulk and $\mathbb{Z}_2$ sectors can be used for model building, as can be seen from the corresponding worldsheet CFT~\cite{Blumenhagen:1999md,Blumenhagen:1999ev,Forste:2000hx,Blumenhagen:2002wn}.

The number of independent conditions on the existence of a $\mathbb{Z}_n$ symmetry is $(h_{21}^{({\rm bulk}+\mathbb{Z}_2)}+1)=6$ for $\mathbb{Z}_6^{(\prime)}$ and 16 for $\mathbb{Z}_2 \times \mathbb{Z}_6^{(\prime)}$. Their exact shape depends on the choice of lattice orientation, which can be reduced to physically inequivalent ones, 
namely {\bf AAA}, {\bf A/BAB} and {\bf BBB} for $\mathbb{Z}_6$, only {\bf AAA} and {\bf BBB} for $\mathbb{Z}_2 \times \mathbb{Z}_6'$,
{\bf a/bAA} and {\bf a/bBB} for $\mathbb{Z}_6'$ and only {\bf a/bAA} for $\mathbb{Z}_2 \times \mathbb{Z}_6$.
Details on the reduction can be found in~\cite{Honecker:2012qr} and~\cite{Ecker:2014hma}, respectively.

Considering now for instance the toroidal orbifold $T^6/\mathbb{Z}_6$ encoded by the shift vector $\vec{v}=\frac{1}{6}(-2,1,1)$ on the {\bf BAA} $\leftrightarrow {\bf BBB}$ lattice configuration, we can demystify the statements surrounding equation (\ref{Eq:def-m-integer}). A fractional three-cycle on $T^6/\mathbb{Z}_6$ can be decomposed with respect to the basis consisting of bulk three-cycles $\rho_{i=1,2}$ and exceptional three-cycles $\varepsilon_\alpha$ and $\tilde \varepsilon_\alpha$ with $\alpha \in \{1, \ldots, 5 \}$ at the $\mathbb{Z}_2\subset \mathbb{Z}_6$ orbifold singularities:
\begin{equation}\label{Eq:Z6-3-cycle-expansion}
\Pi_a = \frac{1}{2} \left( X_a \, \rho_1 + Y_a \, \rho_2 + \sum_{\alpha=1}^5 \left[x_{a,\alpha} \,\varepsilon_{\alpha} + y_{a,\alpha}  \,\tilde{\varepsilon}_{\alpha}  \right] \right)  ,
\end{equation}
where the bulk wrapping numbers $(X_a, Y_a)$ and exceptional wrapping numbers $(x_{a,\alpha}, y_{a, \alpha})$ are all integer-valued. This basis does however not correspond to the basis $(\Pi_{i}^{\rm even}, \Pi_{i}^{\rm odd})_{i\in\{1,\ldots, h_{21}+1\}}$ adapted to the $\Omega {\cal R}$-projection on the {\bf BAA} background lattice, as can be seen from table~\ref{Tab:OR-on-Z6}. 
\begin{table}[h!]
\caption{\label{Tab:OR-on-Z6}Orientifold projection on the bulk three-cycles $\rho_{i \in \{1,2\}}$ and exceptional three-cycles $\varepsilon_{\alpha \in \{1 \ldots 5\}}$, $\tilde{\varepsilon}_{\alpha \in \{1 \ldots 5\}}$
on $T^6/(\mathbb{Z}_6 \times \Omega {\cal R})$ with {\bf BAA} background orientation.}
  \begin{center}
\begin{tabular}{ccc}\br
basis cycle & $\Omega {\cal R}$-image & \\
\hline
$\rho_1$ & $\rho_2$ & \\
$\rho_2$ & $\rho_1$ & \\
$\varepsilon_{\alpha}$ & $- \tilde{\varepsilon}_{\beta}$  & $\alpha=\beta=1,2,3 ; \;  \alpha=4,5  \leftrightarrow \beta=5,4 $  \\
$\tilde{\varepsilon}_{\alpha}$ &   $- \varepsilon_{\beta} $ & $\alpha=\beta=1,2,3 ; \;  \alpha=4,5  \leftrightarrow \beta=5,4 $ \\
\hline
\end{tabular}
\end{center}
\end{table}

\noindent The basis of ${\Omega} {\cal R}$-even and ${\Omega} {\cal R}$-odd three-cycles for the {\bf BAA} lattice configuration reads:
\begin{equation}
\begin{array}{ll}
\Pi^{\rm even}_0 =\rho_1 + \rho_2
,
&
\Pi^{\rm  odd}_0 = \rho_1 - \rho_2
,
\\
\Pi^{\rm even}_{\alpha} = \left\{\begin{array}{c} 
\varepsilon_{\alpha} - \tilde{\varepsilon}_{\alpha}  
\\ \varepsilon_4 - \tilde{\varepsilon}_5 
\\ \varepsilon_5 - \tilde{\varepsilon}_4 
\end{array}\right. 
,
\qquad
&
\Pi^{\rm odd}_{\alpha} = \left\{\begin{array}{cr}  
\varepsilon_{\alpha} + \tilde{\varepsilon}_{\alpha} & \alpha=1,2,3 \\ 
\varepsilon_5 + \tilde{\varepsilon}_4 & 4 \\ \varepsilon_4 + \tilde{\varepsilon}_5 & 5 \end{array}\right. 
,
\end{array}
\end{equation}
where the basis three-cycles now topologically intersect as in equation (\ref{Eq:def-m-integer}) with $m_i=4$. Expressing a generic fractional three-cycle (\ref{Eq:Z6-3-cycle-expansion}) now in terms of the ${\Omega} {\cal R}$-even and ${\Omega} {\cal R}$-odd three-cycles leads to the expansion:
\begin{eqnarray}
\Pi_a &=& \frac{1}{4} \left[ \left(X_a + Y_a \right) \Pi^{\rm even}_0 + \sum_{\alpha=1}^3 \left(x_{a,\alpha} - y_{a,\alpha} \right) \Pi^{\rm even}_\alpha + \sum_{(\alpha, \beta) \in \{(4,5), (5,4) \}}  \left(x_{a,\alpha} - y_{a,\beta} \right) \Pi^{\rm even}_\alpha  \right] \\
&&+ \frac{1}{4} \left[ \left(X_a - Y_a \right) \Pi^{\rm odd}_0 + \sum_{\alpha=1}^3 \left(x_{a,\alpha} + y_{a,\alpha} \right) \Pi^{\rm odd}_\alpha + \sum_{(\alpha, \beta)\in\{(4,5), (5,4)\}}  \left(x_{a,\beta} + y_{a,\alpha} \right) \Pi^{\rm odd}_\alpha   \right] \nonumber.
\end{eqnarray}
By comparing the coefficients in this expansion to the expressions in equation (\ref{Eq:even+odd-cycles-expansion}), one clearly sees that the coefficients $A^i_a$ and $B^i_a$ take rational values in $\frac{1}{4} \mathbb{Z}$.

\subsection{Local versus global D-brane models}\label{Ss:local-global}

In {\it local} models such as the gauge quivers in~\cite{Anastasopoulos:2012zu}, the lattice of three-cycles is not given. 
In~\cite{Anastasopoulos:2012zu}, the postulated chiral particle multiplicities or
intersection numbers per gauge quiver were combined in order to produce {\it necessary} conditions 
on the existence of discrete $\mathbb{Z}_n$ symmetries along the lines of equation~(\ref{Eq:discrete-as-intersection}). To this end, the ${\cal R}$-even combinations 
$\Pi_x + \Pi_x^{\prime}$ of three-cycles were used to compute intersection numbers $(\Pi_x + \Pi_x^{\prime}) \circ \sum_a N_a k_a  \Pi_a$. 
For $U(3)_a \times U(2)_b \times U(1)_c \times U(1)_d$ and $U(3)_a \times USp(2)_b \times U(1)_c$ gauge quivers, such intersection numbers with $x\in \{a,b,c,d\}$ 
and $x \in \{a,c\}$, respectively, provide at most four or two  necessary conditions on the existence of $\mathbb{Z}_n$ symmetries. 
These conditions are not {\it sufficient} since first of all $(h_{21}^{({\rm bulk} + \mathbb{Z}_2)}+1)$ is generically greater than four, e.g. 6 or 16 in the examples of section~\ref{Ss:Orbifolds}, 
and secondly it is not guaranteed that the above intersection numbers correspond to $\mathcal{R}$-even cycles of minimal length.

Last but not least, the mixing of Abelian gauge bosons from different stacks of D-branes is an intrinsic feature of {\it global} models. Any embedding of the {\it local} gauge quivers into a D-brane set-up satisfying RR tadpole cancellation and K-theory constraints might thus potentially radically change the structure of discrete Abelian symmetries.

\subsection{A generation-dependent discrete symmetry}\label{Ss:generation-dependent}

In~\cite{Honecker:2012qr}, a Pati-Salam model with three particle generations on the orbifold $T^6/\mathbb{Z}_2 \times \mathbb{Z}_6'$ with discrete torsion ($\eta=-1$) was constructed.
The initial gauge group is $U(4)_a \times U(2)_b \times U(2)_c \times U(2)_d \times U(2)_e$, and all five Abelian factors acquire a mass at the string scale.
The massless open string spectrum consists of the `chiral' part $[C]$ obtained from non-trivial topological intersection numbers and vector-like states $[V]$ that are obtained by using either the method of Chan-Paton factors or  the beta function coefficients given in~\cite{Honecker:2011sm},
\begin{eqnarray}\label{Eq:Spectrum-ex1}
{} [C] 
&=&
 ({\bf 4},\overline{{\bf 2}},{\bf 1};{\bf 1},{\bf 1}) + 2 \times ( {\bf 4},{\bf 2},{\bf 1};{\bf 1},{\bf 1}) + ( \overline{{\bf 4}},{\bf 1},{\bf 2};{\bf 1},{\bf 1}) 
+ 2 \times ( \overline{{\bf 4}},{\bf 1},\overline{{\bf 2}};{\bf 1},{\bf 1}) + ( {\bf 1},{\bf 2},\overline{{\bf 2}};{\bf 1},{\bf 1})
\nonumber
\\
&+& ({\bf 1},{\bf 2},{\bf 1};\overline{{\bf 2}},{\bf 1}) + 3 ({\bf 1},\overline{{\bf 2}},{\bf 1};\overline{{\bf 2}},{\bf 1}) + ({\bf 1},\overline{{\bf 2}},{\bf 1};{\bf 1},\overline{{\bf 2}}) 
+ ({\bf 1},{\bf 1},\overline{{\bf 2}};{\bf 2},{\bf 1}) + 3 ({\bf 1},{\bf 1},{\bf 2};{\bf 2},{\bf 1}) + ({\bf 1},{\bf 1},{\bf 2};{\bf 1},{\bf 2})
\nonumber
\\
& \equiv & (Q_L,L)_{ab} + 2 \times (Q_L,L)_{ab'} + (Q_R,R)_{ac} + 2 \times  (Q_R,R)_{ac'} + (H_d,H_u)
\\
&+& X_{bd} + 3 \times X_{bd'} + X_{be'} + X_{cd} + 3 \times X_{cd'} + X_{ce'}
,\nonumber
\\
{}[V]
&=&  2 \times \left[({\bf 4},{\bf 1},{\bf 1}; \overline{{\bf 2}},{\bf 1}) + h.c. \right] + \left[({\bf 1},{\bf 1},{\bf 1};{\bf 2},{\bf 2}) + h.c. \right]  +  ({\bf 1},{\bf 1},{\bf 1}; {\bf 4}_{\rm Adj},{\bf 1}) 
\nonumber\\
&+  &  2 \times \left[({\bf 1},{\bf 1},{\bf 1};{\bf 3}_{\bf S},{\bf 1}) + ({\bf 1},{\bf 1},{\bf 1};{\bf 1}_{\bf A},{\bf 1}) + h.c. \right] + \left[ ({\bf 1},{\bf 1},{\bf 1};{\bf 1},{\bf 3}_{\bf S}) + ({\bf 1},{\bf 1},{\bf 1};{\bf 1},{\bf 1}_{\bf A}) + h.c. \right]
.\nonumber
\end{eqnarray}
The number of constraint equations~(\ref{Eq:discrete-as-intersection}) on discrete $\mathbb{Z}_n$ symmetries is given by $(h_{21}+1)=16$ in this example with at most five linearly independent solutions due to the five massive Abelian gauge bosons. It turns out that besides the field theoretically trivial solutions $(k_a,k_b,k_c,k_d,k_e)=(\underline{1,0,0,0,0})$ for $n=4,2,2 \ldots$, there exists one non-trivial solution $(0,1,1,1,1)$ for $n=4$~\cite{Honecker:2013hda}. The corresponding charges of particles in part $[C]$ of the massless spectrum are listed in table~\ref{Tab:PS-GenerationDependent}.
\begin{table}[h!]
\caption{\label{Tab:PS-GenerationDependent} Charge assignment of the generation-dependent $\mathbb{Z}_4$  symmetry on $T^6/\mathbb{Z}_2 \times \mathbb{Z}_6'$.
The charge assignments of the reduced $\mathbb{Z}_2$ version are obtained from inspections of 3-point couplings.}
  \begin{center}
\begin{tabular}{ccccccccccccc}\br
& \multicolumn{2}{c}{$\begin{array}{c}(Q_L,L)\\ ab \hspace{0.2in}  ab' \end{array}$} &  \multicolumn{2}{c}{ $\begin{array}{c}(Q_R,R)\\ ac \hspace{0.2in} ac' \end{array}$} & $(H_d,H_u)$  &$X_{bd}$ & $X_{bd'}$ & $X_{be'}$ & $X_{cd}$ & $X_{ cd'}$& $X_{ce'}$ 
\\\mr
$\mathbb{Z}_4$ & 3 & 1 & 1&  3 & 0 & 0 & 2 & 2 & 0 & 2 & 2
\\\mr
$\mathbb{Z}_2$ & 0 & 1 & 0 & 1 & 0 & 0 & 1 & 1 & 0 & 1 & 1
\\\br
\end{tabular}
\end{center}
\end{table}

There exists the unwritten folklore that ultimately the non-trivial Abelian discrete symmetries in field theory are obtained by modding out the $\mathbb{Z}_N \subset U(N)$ symmetries, which in this example amounts to $((\mathbb{Z}_4)^2 \times (\mathbb{Z}_2)^3)/(\mathbb{Z}_4 \times (\mathbb{Z}_2)^4) \simeq \mathbb{Z}_2$. However, the charge assignments for this $\mathbb{Z}_2$ symmetry cannot be obtained by simply taking `mod 2' of the generation-dependent $\mathbb{Z}_4$ charges.
For example, one can consider the following three-point couplings:
\begin{itemize}
\item $(Q_L,L)_{ab} .(Q_R,R)_{ac}.(H_d,H_u)$ is perturbatively allowed,
\item $(Q_L,L)_{ab} .(\overline{\bf 4},{\bf 1},{\bf 1},{\bf 2},{\bf 1}). ({\bf 1},\overline{\bf 2},{\bf 1},\overline{\bf 2},{\bf 1})$ is perturbatively forbidden by the global $U(1)_b$ symmetry
and non-perturbatively constrained by $\mathbb{Z}_4$ charge `2 mod 4',
\item $(Q_L,L)_{ab} .(Q_R,R)_{ac'}. (H_d,H_u)$  is perturbatively forbidden by $U(1)_c$ and has $\mathbb{Z}_4$ charge `2 mod 4'.
\end{itemize}
An extensive scan of three-point couplings involving particles from both sectors $[C]$ and $[V]$ reveals that one can reduce the $\mathbb{Z}_4$ charges (0,2) to $\mathbb{Z}_2$ charges (0,1), but with this choice no clear reduction of $\mathbb{Z}_4$ charges (1,3) to $\mathbb{Z}_2$ charges is possible. The same holds true if instead first the mapping of the charges (1,3) is considered.
Thus, while the reduction to a non-trivial $\mathbb{Z}_2$ group can be done `by hand' and might be interesting for comparison with purely field theoretical models, only the $\mathbb{Z}_4$ symmetry can serve as guide to the ultimate non-perturbative $m$-point coupling selection rules.

\subsection{Discrete symmetries in models with axion candidates}\label{Ss:models+axions}

In the following, we briefly discuss two types of gobal models which contain open string axion candidates used in the discussion of section~\ref{Ss:TypeII-axions}. In the first example in section~\ref{Sss:T6-Z6}, all discrete $\mathbb{Z}_n$ symmetries are trivial from the low-energy effective field theory point of view, while in the second example in section~\ref{Sss:T6-Z6p}, there exists a non-trivial $\mathbb{Z}_3$ symmetry under which the axion is charged.

\subsubsection{A left-right symmetric model on $T^6/\mathbb{Z}_6$:}\label{Sss:T6-Z6}

In~\cite{Honecker:2004kb} a left-right symmetric three-generational model with initial gauge group $U(3)_a\times U(2)_b \times USp(2)_c\times U(1)_d \times USp(2)_e$ was found.  The three Abelian gauge factors combine to form a massless $U(1)_{B-L} = (\frac{Q_a}{3} + Q_d)_{\rm massless}$ symmetry, while the remaining two Abelian factors $U(1)_{\rm massive}^2$ acquire masses at the string scale. 
The massless open string spectrum consists of a part $[C]$ determined by non-vanishing intersection numbers among cycles and a part $[V]$ which is not captured by these topological numbers, but can be computed using Chan-Paton labels or the beta function coefficients as done in~\cite{Gmeiner:2009fb}:
\begin{eqnarray}\label{Eq:Spectrum-ex2}
{} [C] 
&=& 3\times \bigg[
    \left({\bf 3},{\bf 2},{\bf 1} \right)^{(0)}_{\bf 1/3} + \left(\overline{\bf 3},{\bf 1},{\bf 2} \right)^{(0)}_{\bf -1/3}
  + \left({\bf 1},{\bf 1},{\bf 2} \right)^{(1)}_{\bf 1}
 + \left({\bf 1},\overline{\bf 2},{\bf 1} \right)_{\bf -1}^{(-1)} + \left({\bf 1},\overline{\bf 2}, {\bf 1};{\bf 2} \right)_{\bf 0}^{(0)}
\bigg] \nonumber
\\
&  \equiv&  3 \times \bigg[Q_L + (d_R,u_R) + (\nu_R,e_R) + L +\tilde{H}  \bigg]
,
\\
{} [V] 
&=& 
 4 \times \left(\bf{8},{\bf 1},{\bf 1} \right)_{\bf 0}^{(0)}
  + 4 \times \left({\bf 1},{\bf 3},{\bf 1} \right)_{\bf 0}^{(0)}
  +  \left({\bf 1},{\bf 1},{\bf 3} \right)_{\bf 0}^{(0)}
  + (4_a+4_b+4_d)  \times \left({\bf 1},{\bf 1},{\bf 1}\right)_{\bf 0}^{(0)}
 + ({\bf 1},{\bf 1},{\bf 1};{\bf 3})_{\bf 0}^{(0)}
 \nonumber
\\
&+&  \bigg[
  1_m \times   \left({\bf 3},{\bf 2},{\bf 1} \right)_{\bf 1/3}^{(0)}
+3 \times     \left({\bf 3},\overline{\bf 2}, {\bf 1} \right)_{\bf 1/3}^{(0)}
+  1_m \times  \left(\overline{\bf 3},{\bf 1},{\bf 2} \right)_{\bf -1/3}^{(0)}
 + 2_m \times \left(\overline{\bf 3},{\bf 1},{\bf 1} \right)_{\bf 2/3}^{(1)}
\nonumber
\\
&& \quad
   + 3 \times \left(\overline{\bf 3},{\bf 1},{\bf 1} \right)_{\bf -4/3}^{(-1)}
   +  1_m \times  \left({\bf 1},{\bf 2},{\bf 2}\right)_{\bf 0}^{(0)}
+ 1_m \times \left(\overline{\bf 3},{\bf 1},{\bf 1};{\bf 2} \right)_{\bf -1/3}^{(0)}
+  1_m \times \left({\bf 1},{\bf 2},{\bf 1};{\bf 2} \right)_{\bf 0}^{(0)}
\nonumber
\\
&& \quad
 + \left(3+1_m  \right) \times (\overline{\bf 3}_A,{\bf 1},{\bf 1})_{\bf 2/3}^{(0)}
  + \left(3+ 1_m\right) \times \underbrace{({\bf 1},{\bf 1}_A,{\bf 1})_{\bf 0}^{(0)}}
   + \;h.c.\;\bigg]
   ,    \nonumber
   \\
   & &\hspace{78mm} \equiv \Sigma_b \label{Eq:VectorT6Z6}
\end{eqnarray}
where the lower index denotes the $(B-L)$ charge and the upper index in parenthesis the $U(1)_d$ charge assignment.

In this example, the right-symmetric and `hidden' gauge groups arise from D6-branes parallel to the O6-planes along the $\mathbb{Z}_2$ invariant two-torus. Therefore, the gauge groups can be broken  $USp(2)_{x \in \{c,e\}} \to U(1)_{x,{\rm massless}}$ by brane displacement away from the O6-plane position, and the hypercharge is obtained as the `standard' realization $Q_Y =\frac{Q_a}{6} + \frac{Q_c+Q_d}{2}$.

Even before the breaking of $USp(2)_c$, the massless $(B-L)$ charge can be used to eliminate $\mathbb{Z}_n$ symmetries that are trivial in field theory. 
In this example, $(k_a,k_b,k_d)=(1,0,1)$ and $(0,1,0)$ solve the constraint equations~(\ref{Eq:discrete-as-intersection}) for $n=2$, and $(1,0,0)$ for $n=3$. At most two discrete symmetries can be independent since they originate from two massive Abelian gauge bosons. In this example, however, all discrete $\mathbb{Z}_n$ symmetries are trivial from the low-energy effective field theory point of view, in case of $(1,0,1)$ upon using the $(B-L)$-shift and for $(1,0,0)$ and $(0,1,0)$ due to their appearance as $\mathbb{Z}_N \subset U(N)$.

For later reference in section~\ref{S:OpenAxions}, the open string axion candidate $\Sigma_b$ is highlighted in the massless spectrum.

\subsubsection{A left-right symmetric model on $T^6/\mathbb{Z}_6^{\prime}$:}\label{Sss:T6-Z6p}

In~\cite{Gmeiner:2007zz,Gmeiner:2008xq}, left-right symmetric three-generational models on the orbifold $T^6/\mathbb{Z}_6^{\prime}$ were found without and with `hidden' gauge group. Starting from $U(3)_a\times U(2)_b\times USp(2)_c \times U(1)_d \; (\times USp(6)_{\rm hidden})$, as in the previous example the Abelian factors split into 
$U(1)_{B-L} = (\frac{Q_a}{3} + Q_d)_{\rm massless}$ and $U(1)_{\rm massive}^2$. The right-symmetric group $USp(2)_c$ can also in this example be broken by a brane displacement. But in the example with `hidden' sector, the corresponding stack of D6-branes is perpendicular to the O6-planes along the $\mathbb{Z}_2$ invariant direction and can thus not be broken by a continuous displacement or Wilson line. Its phenomenologically appealing feature consists in the fact that the amount of matter charged under the `hidden' gauge group is small enough such that its gauge coupling runs to the strong regime below the string scale, and supersymmetry might be broken by the formation of a gaugino condensate. 

The massless open string spectrum has now two or three components (with $\omega=1,2$ for the model with and without `hidden' sector, respectively),
\begin{eqnarray}\label{Eq:Spectrum-ex3}
{} [C] 
&=& 3\times \bigg[
    \left({\bf 3},{\bf 2},{\bf 1} \right)^{(0)}_{\bf 1/3} + \left(\overline{\bf 3},{\bf 1},{\bf 2} \right)^{(0)}_{\bf -1/3}  + \left({\bf 1},{\bf 1},{\bf 2} \right)^{(1)}_{\bf 1}
 + 2 \times \left({\bf 1},{\bf 2},{\bf 1} \right)_{\bf -1}^{(-1)} +  \left({\bf 1},{\bf 2},{\bf 1} \right)_{\bf 1}^{(1)} 
\bigg] \nonumber
\\
& +& 9 \times \biggl[ \omega \times \left({\bf 1},\overline{\bf 2},{\bf 2} \right)_{\bf 0}^{(0)} 
+ (\omega -1) \times  \left({\bf 1},{\bf 1}_{\overline{A}},{\bf 1} \right)_{\bf 0}^{(0)} 
\biggr]
\\
&  \equiv&  3 \times \bigg[Q_L + (d_R,u_R) + (\nu_R,e_R) + 2 \times L +\overline{L}  \bigg]
+ 9 \times \bigg[ \omega \times (H_d,H_u) + (\omega -1 )\times \overline{\Sigma}_b \bigg]
, \nonumber 
\\
{} [V_U] 
&=&
2 \times \left(\bf{8},{\bf 1},{\bf 1}\right)_{\bf 0}^{(0)}
  + 10 \times \left({\bf 1},{\bf 3},{\bf 1}\right)_{\bf 0}^{(0)}
+ \left({\bf 1},{\bf 1},{\bf 3} \right)_{\bf 0}^{(0)}
  + (2_a+10_b +3_c +10_d )  \times \left({\bf 1},{\bf 1}, {\bf 1}\right)_{\bf 0}^{(0)}
  \nonumber
\\
& +&  \bigg[
     \left({\bf 3},{\bf 2},{\bf 1}\right)_{\bf 1/3}^{(0)}
  + 3\times \left(\overline{{\bf 3}},{\bf 1},{\bf 1}\right)_{\bf 2/3}^{(1)}
  + 3 \times \left(\overline{{\bf 3}},{\bf 1},{\bf 1}\right)_{\bf -4/3}^{(-1)}
   + 2 \times (\overline{{\bf 3}}_A,{\bf 1},{\bf 1})_{\bf 2/3}^{(0)} + ({\bf 6}_S,{\bf 1},{\bf 1})_{\bf 2/3}^{(0)} 
   \nonumber
\\
&&\quad
  + \omega_m \times  \left({\bf 1},\overline{\bf 2},{\bf 2}\right)_{\bf 0}^{(0)}
  + 2_m \times  \left({\bf 1},{\bf 2},{\bf 1}\right)_{\bf -1}^{(-1)}
  + 1_m \times  \left({\bf 1},{\bf 2},{\bf 1}\right)_{\bf 1}^{(1)}
  + 1_m \times   ({\bf 1},{\bf 1},{\bf 2})_{\bf 1}^{(1)}
  \nonumber\\
&&\quad
  + \left( 3 \,( \omega+1) + (\omega -1)_m \right) \times ({\bf 1},{\bf 3}_S,{\bf 1})_{\bf 0}^{(0)}
 + \left( 2 +2 \, \omega \right)_m \times \underbrace{({\bf 1},{\bf 1}_{A},{\bf 1})_{\bf 0}^{(0)}}
  + \;h.c.\;\bigg]
  ,
  \nonumber
  \\
  && \hspace{100mm} \equiv \Sigma_b  \label{Eq:VectorT6Z6p}
\\
{} [V_{h_3}] &=&
2 \times ({\bf 1},{\bf 1},{\bf 1};{\bf 15})^{(0)}_{\bf 0} + 2 \times ({\bf 1},{\bf 1},{\bf 2};{\bf 6})_{0}^{(0)}
+ \bigg[  ({\bf 1},{\bf 2},{\bf 1};{\bf 6})^{(0)}_{\bf 0}  \; + h.c. \; \bigg] 
,
\end{eqnarray}
where corrections of multiplicities in $[V_U]+[V_{h_3}]$ have been taken into account, cf.~\cite{Honecker:2012jd} for the model with hidden $USp(6)_{\rm hidden}$ and~\cite{Forste:2010gw,Honecker:2012qr} for details on the sign factors in the counting of multiplicities of (anti)symmetric representations.

The constraint equations~(\ref{Eq:discrete-as-intersection}) are solved by $(k_a,k_b,k_d) = (1,0,1)$ for $n=2$, by $(1,0,0)$ for $n=3$ and by $(0,1,0)$ for $n=6$, both for the model without as well as for the model with `hidden' sector. A shift over the massless $(B-L)$ symmetry can be used to eliminate the discrete symmetries acting trivially in field theory, namely the $\mathbb{Z}_2$ and $\mathbb{Z}_3$ discrete symmetries. The non-trivial $\mathbb{Z}_6$ discrete symmetry reduces to a $\mathbb{Z}_3$ symmetry by an additional hypercharge shift upon the spontaneous breaking of $USp(2)_c$ to a massless $U(1)_c$ as described above. The charges of Standard Model particles and axion candidates are displayed in table~\ref{Tab:Z3-charges-of-Z6p}, where after $U(1)_c$ rotation `mod 2' of the original charge assignments has been taken, as the charges after the $U(1)_c$ rotation only took values in $\{0,2,4\}$.
\begin{table}[h!]
\caption{\label{Tab:Z3-charges-of-Z6p} Charge assignments of the non-trivial $\mathbb{Z}_3$ symmetry in the $T^6/\mathbb{Z}_6'$ model.}
 \begin{center}
\begin{tabular}{ccccccccccccc}\br
 & $Q_L$ & \multicolumn{2}{c}{$(u_R,d_R)$} & $L$ & $\overline{L}$ & \multicolumn{2}{c}{$(e_R,\nu_R)$} &  \multicolumn{2}{c}{$(H_u,H_d)$} & $\Sigma_b$
\\\mr
$\mathbb{Z}_6$ & 0 &  \multicolumn{2}{c}{1} & 4 & 4 &  \multicolumn{2}{c}{3} &   \multicolumn{2}{c}{5} & 2
\\\mr
$\stackrel{U(1)_c, {\rm mod} \, 2}{\longrightarrow} \mathbb{Z}_3$ & 0 & \multicolumn{2}{c}{0 \quad 1}  & 2 & 2 & \multicolumn{2}{c}{2 \quad1}  & \multicolumn{2}{c}{0 \quad 2}  & 1
\\\br
\end{tabular}
\end{center}
\end{table}

Both the discrete $\mathbb{Z}_6$ symmetry and its reduced version as $\mathbb{Z}_3$ symmetry have not yet appeared before in the literature,
and the non-trivial charge assignment of the axion $\Sigma_b$ provides the ultimate non-perturbative selection rule on its couplings beyond the perturbative global $U(1)_b \simeq U(1)_{\rm PQ}$ symmetry.

\section{Open string axions and axion mixing}\label{S:OpenAxions}

In this section, we justify the identification of field theoretical axion candidates $\Sigma_b$ in the models of section~\ref{Ss:models+axions}
and then discuss the mixing among open and closed string axions in section~\ref{Ss:Open-closed-mixing}.

\subsection{Open string axions in Type II string compactifications}\label{Ss:TypeII-axions}

Axions were originally invoked to solve the strong CP-problem by coupling the pseudoscalar $\alpha$ to the field strength $G_{\mu\nu}$ of the strong interaction,
\begin{equation}
\mathcal{L}_{\alpha} \supset \frac{1}{2} \left(\partial_\mu \alpha \right) \left(\partial^\mu \alpha \right) -  \frac{1}{32 \pi^2} \frac{\alpha(x)}{f_{\alpha}}\, {\rm Tr} ( G_{ \mu \nu} \tilde G^{\mu \nu}) ,
\end{equation}
where $f_{\alpha}$ is the axion decay constant. 
In the original model~\cite{Peccei:1977hh,Peccei:1977ur,Weinberg:1977ma,Wilczek:1977pj}, the electro-weak and Peccei-Quinn breaking scales are identical and thus experimentally excluded by e.g. the so far non-observation of axion/photon conversion in astrophysical and laboratory searches.

The two symmetry breaking scales can be decoupled by introducing a new scalar Standard Model singlet field $\sigma$ which carries $U(1)_{\rm PQ}$ charge and modifies the Higgs potential. 
The DSFZ model~\cite{Dine:1981rt,Zhitnitsky:1980tq} already contains two Higgs doublets and can be easily promoted to a supersymmetric version with $\Sigma$ the chiral multiplet associated to the scalar $\sigma$. However, the embedding within Type II string theory models is severely constraining since each open string has two endpoints which each transform in the (anti)fundamental representation of the gauge group supported by the stack of D-branes to which the open string end point is attached. In case the two stacks of D-branes are $\mathcal{R}$-images of each other, the bifundamental representation is replaced by the antisymmetric and/or symmetric representation of the single gauge factor. 

For a massive Abelian symmetry to act as global Peccei-Quinn symmetry $U(1)_{\rm PQ}$ in the perturbative low-energy effective field theory and the `standard' ansatz of embedding the Standard Model in a $U(3)_a \times U(2)_b \times U(1)_c \times U(1)_d \times G_{\rm hidden}$ quiver gauge theory, we have to demand the following~\cite{Honecker:2013mya}:
\begin{itemize}
\item
the scalar $\sigma$ does not carry $SU(3)_a \times SU(2)_b$ charge, i.e.~the corresponding open string cannot have any endpoint on the $U(3)_a$-stack; the scalar $\sigma$ can either transform in the antisymmetric representation of $U(2)_b$ or not carry any $U(1)_b$ charge at all;
\item
the Higgs doublets $(H_d,H_u)$ are charged under $U(1)_{\rm PQ}$; since the Higgs fields arise at some intersections of the type $\{(bc),(bc'),(bd),(bd')\}$, either $U(1)_{b}$ or the corresponding $U(1)_{x, x \in \{c,d\}}$ groups  (possibly with admixtures from Abelian factors within $G_{\rm hidden}$) or some linear combination thereof form $U(1)_{\rm PQ}$;
\item
$\sigma$ does not carry hypercharge; supposing the `standard' definition $Q_Y=\frac{Q_a}{6} + \frac{Q_c+Q_d}{2}$, the Peccei-Quinn symmetry is either generated by $U(1)_b$ or $U(1)_c - U(1)_d$ or superpositions of the two (and possibly factors within $G_{\rm hidden}$);
\item
the $U(1)_{\rm PQ}$ charge assignment is compatible with non-vanishing quark and lepton Yukawa couplings as well as with cross-terms between $(H_d,H_u)$ and $\sigma$ in the scalar potential.
\end{itemize}
Within the low-energy effective field theory, the massless hypercharge can be used to ensure that either $(Q_L)$ or $(d_R,u_R)$ have vanishing $U(1)_{\rm PQ}$ charge.
Following the above reasoning for an embedding into Type II string theory, the first choice corresponds to $U(1)_{\rm PQ}=U(1)_c - U(1)_d$ with some right-handed sneutrino as Standard Model singlet $\sigma$,  and the second choice corresponds to $U(1)_{\rm PQ}=U(1)_b$ with $\sigma^{(\ast)} \simeq ({\bf Anti}_{U(2)_b})$. 
In both cases, $\sigma$ carries (up to a sign) twice the $U(1)_{\rm PQ}$ charge of $(H_d,H_u)$ as summarized in table~\ref{Tab:PQ-chages}.
\begin{table}[h!]
\caption{\label{Tab:PQ-chages} $U(1)_{\rm PQ} \simeq U(1)_b$ charge assignments for the stringy DFSZ axion model.}
  \begin{center}
\begin{tabular}{ccccccccccccc}\br
Matter & $Q_L$ & $u_R$ & $d_R$ & $H_u$ & $H_d$ & $L$ & $e_R$ & $\nu_R$ & $\Sigma$
\\\mr
$q_{PQ}$ & $ \mp 1 $ & $ 0 $ & $ 0 $ &$\pm 1$ &  $\pm 1$  &$\mp 1$& $0$& $0$&  $\mp2$
\\\br
\end{tabular}
\end{center}
\end{table}

The stringy version of the Higgs-axion potential in the supersymmetrized DFSZ model~\cite{Rajagopal:1990yx} reads,
\begin{eqnarray}
V^{\rm stringy}_{\rm DFSZ}(H_u, H_d, \sigma) &=& \lambda_u ( \underbrace{H_u^\dagger H_u} -   v_u^2)^2 
+ \lambda_d (  \underbrace{H_d^\dagger H_d} -  v_d^2 )^2  +  \lambda_\sigma (   \underbrace{\sigma^* \sigma} -   v_\sigma^2)^2
+ \underbrace{ a\, |  H_u^\dagger H_d |^2} \nonumber
 \\
 && \hspace{7mm} \subset V_D  \hspace{21mm} \subset V_D  \hspace{20mm} \subset V_D  \hspace{17mm} \subset V_D 
\nonumber \\
&& + \underbrace{(b_1\, H_u^\dagger H_u + b_2\, H_d^\dagger H_d )\, \sigma^* \sigma  + b_3\, | H_u \cdot H_d |^2}
+ ( c\,  H_u \cdot H_d\, \sigma  + h.c. ),
\nonumber \\
 && \hspace{30mm} \subset V_F
\end{eqnarray}
where in contrast to the original DFSZ model, the last line contains the coupling $H_u \cdot H_d\, \sigma$ with $\sigma$ appearing only linearly instead of quadratically.
This is due to the doubled charge in table~\ref{Tab:PQ-chages} of the axion supermultiplet $\Sigma$ in Type II string compactifications. 
All terms containing $(H_d,H_u)$ or $\sigma$ on the first line can be deduced from D-terms, and the underbraced terms in the second line can be traced back to F-terms
originating from the superpotential $\mathcal{W}= \mu \Sigma H_d \cdot H_u$, implying $b_1 = b_2 = b_3 = |\mu|^2$. 
The remaining terms can only originate from soft supersymmetry breaking.

If in addition, we demand that the Peccei-Quinn symmetry does not mix with the gauged $(B-L)$ symmetry in the (to the Standard Model broken phase of the) $T^6/\mathbb{Z}_6$ and  $T^6/\mathbb{Z}_6^{\prime}$ models of section~\ref{Ss:models+axions}, the choice $U(1)_c - U(1)_d$ is ruled out, and we have the unique option of identifying $U(1)_{\rm PQ} \simeq U(1)_b$, and $\Sigma$ is one of the states $\Sigma_b$ from equation~(\ref{Eq:VectorT6Z6}) and (\ref{Eq:VectorT6Z6p}), respectively.

\subsection{Mixing of open and closed string axions}\label{Ss:Open-closed-mixing}

The physical open string axion $a$ arises as the complex phase of the Standard Model singlet scalar $\sigma$,
\begin{eqnarray}
\sigma = \frac{v_{\sigma}+s(x)}{\sqrt{2}} e^{i\, \frac{a(x)}{v_{\sigma}}},
\end{eqnarray}
which mixes with the closed string axion $\xi$ to form the massless axion $\alpha$ and the massive axion $\zeta$,
\begin{equation}
\zeta = \frac{ {\xi} +  \frac{q v_{\sigma}}{M_{\rm string}} \, a}{\sqrt{1  + \left(\frac{q v_{\sigma}}{M_{\rm string}}\right)^2}}, 
\qquad  \alpha =  \frac{ a  -   \frac{q v_{\sigma}}{M_{\rm string}}\, {\xi}}{\sqrt{1 +  \left(\frac{q v_{\sigma}}{M_{\rm string}}\right)^2}},
\end{equation}
with Peccei-Quinn charge $q=2$ as argued in section~\ref{Ss:TypeII-axions} and the mass given by
\begin{eqnarray}\label{Eq:StuckClosedOpenAxion}
{\cal L}_{\rm CP-odd} \supset \frac{1}{2} \left( \partial_\mu \zeta + m_B B_\mu   \right)^2 + \frac{1}{2} (\partial_\mu \alpha)^2
\qquad 
{\rm with}
\qquad
m_B^2 =  M_{\rm string}^2 + q^2 v_{\sigma}^2.
\end{eqnarray}
The two physical axions $(\alpha,\zeta)$ couple to the field strength of the strong interactions via
\begin{eqnarray}
{\cal L}_{\rm anom} =-\frac{1}{32 \pi^2} \frac{\zeta (x)}{ f_\zeta}\Tr (G_{\mu \nu} \tilde G^{\mu \nu})  
- \frac{1}{32 \pi^2} \frac{\alpha (x)}{f_{\alpha}}  \Tr (G_{\mu \nu} \tilde G^{\mu \nu}),
\end{eqnarray}
with the axion decay constants given by 
\begin{equation}
f_{\zeta} = \frac{ M_{\rm string} }{2}  \, \sqrt{1 + \left( \frac{qv_{\sigma}}{M_{\rm string}}\right)^2 }, 
\qquad 
f_{\alpha} = q v_{\sigma} \, \frac{ \sqrt{ 1+ \left( \frac{qv_{\sigma}}{M_{\rm string}}\right)^2 }}{ 1 - \left( \frac{qv_{\sigma}}{M_{\rm string}}\right)^2 }.
\end{equation}
As the axion $\zeta$ is turned into the longitudinal component of the massive $U(1)$ gauge boson through the St\"uckelberg mechanism in equation~(\ref{Eq:StuckClosedOpenAxion}), the candidate for the QCD axion is formed by the orthogonal direction $\alpha$.  In the range where the low-energy effective field theory approach to the breaking of $U(1)_{\rm PQ}$ is reliable, $qv_{\sigma} \ll M_{\rm string}$, the massive axion $\zeta$ consists mainly of the closed string axion $\xi$ and has a decay constant $f_{\zeta} \sim \frac{ M_{\rm string} }{2}$, and the massless axion $\alpha$ consists mostly of the open string axion $a$ with decay constant $f_{\alpha} \sim q v_{\sigma}$. The axion $\alpha$ then serves as the QCD axion whenever $f_{\alpha}$ lies within the axion window, i.e.~$10^{9}$ GeV$\leq f_{\alpha}\leq10^{12}$ GeV. The presence of the open string axion alleviates the constraint on the string scale $M_{\rm string}$, which would otherwise have to lie within the axion window in case the closed string axion would take on the r\^ole of the QCD axion. Lower bounds on $M_{\rm string}$ can also be derived~\cite{Honecker:2013mya} by ensuring the reality of the gauge coupling for a strongly coupled field theory upon taking into account one-loop corrections. String mass scales of the order $M_{\rm string} \sim 10^{9}$~GeV  
are in itself perfectly compatible with this requirement, yet they would force $f_{\alpha}$ to lie outside the axion window. Hence, the string mass scale would favourably lie above $10^{12}$ Gev in order for the QCD $\theta$-problem to be solved by the Peccei-Quinn mechanism in the presented set-up. 

Kinetic mixing effects among multiple axions hide other interesting properties as well, with possible applications to inflationary scenarios with axions~\cite{Shiu:2015xy}.

\section{Conclusions}\label{S:Conclusions}
Massive $U(1)$ gauge symmetries arise naturally in the context of D-brane models as a consequence of the St\"uckelberg couplings inherent to the Green-Schwarz mechanism. The perturbative behaviour at low energy of such a massive $U(1)$ corresponds to a global continuous symmetry, which is expected to be broken by non-perturbative effects such as instantons. The observation that a discrete $\mathbb{Z}_n$ subgroup of the $U(1)$ symmetry may be left unbroken by the instanton effects motivates their usage to constrain $m$-point couplings between massless states beyond perturbation theory. 

In this proceedings article, we briefly reviewed the conditions on the existence of discrete $\mathbb{Z}_n$ symmetries for global intersecting D6-brane models on toroidal orbifolds and Calabi-Yau manifolds, for which the unimodular basis of three-cycles is (at least partially) not aligned with the symmetry planes of the anti-holomorphic orientifold involution $\Omega\mathcal{R}$. Explicit examples of such background lattices can be found on toroidal orbifolds of the type $T^6/\mathbb{Z}_{2N}$ and $T^6/\mathbb{Z}_{2}\times \mathbb{Z}_{2M}$ with discrete torsion, for which the orbifold action implies the presence of $\mathbb{Z}_2$ twisted sectors. 
Moreover, it is also well known that these backgrounds allow for global intersecting D6-branes models, in which case an exhaustive search for discrete $\mathbb{Z}_n$ symmetries is well justified and feasible. For the global five-stack Pati-Salam model on $T^6/\mathbb{Z}_2\times\mathbb{Z}_6'$ with discrete torsion in section~\ref{Ss:generation-dependent}, such an exhaustive search revealed the presence of a generation-dependent $\mathbb{Z}_4$ symmetry, which reduces to a generation-dependent $\mathbb{Z}_2$ symmetry upon modding out the trivial $\mathbb{Z}_N \subset U(N)$ symmetries. Furthermore, in section~\ref{Sss:T6-Z6p} a non-trivial discrete $\mathbb{Z}_3$ symmetry was identified in the left-right symmetric models with and without hidden sector on the $T^6/\mathbb{Z}_6'$ orbifold.

Studying the global $U(1)$ symmetry in the unbroken phase opens up the perspective of identifying it as a Peccei-Quinn symmetry. The presence of two Higgs doublets in the DFSZ axion model facilitates a straightforward generalization to a supersymmetric version realized in the framework of D-brane models. Combining the 
field content of the DFSZ axion model with the `standard' realization of the hypercharge within the Standard Model gauge group on  four intersecting D6-brane stacks only leaves two global $U(1)_{\rm massive}$ symmetries as plausible candidates for a Peccei-Quinn symmetry. Excluding mixing between the Standard Model hypercharge and the Peccei-Quinn symmetry  selects a unique candidate, namely  $U(1)_{\rm PQ} \simeq U(1)_b \subset U(2)_b$ associated to the left-symmetric group $SU(2)_L \simeq SU(2)_b \subset U(2)_b$. The Standard Model singlet field coupling to the Higgses is then identified as a massless state in the antisymmetric representation of $U(2)_b$. Explicit realizations of this construction are given by the left-right symmetric models on $T^6/\mathbb{Z}_6$ and $T^6/\mathbb{Z}_6'$, which we discussed in the context of discrete symmetries in section~\ref{Ss:models+axions}, upon spontaneous breaking of the right-symmetric $SU(2)_R \simeq USp(2)_c\to U(1)_c $ gauge symmetry. The QCD axion in this set-up is identified as a linear combination of a closed string and an open string axion, perpendicular to the axionic direction participating in the St\"uckelberg mechanism. For a sufficiently large string scale $M_{\rm string}$, the open string axion forms the largest portion of the QCD axion, imposing the open string saxion {\it vev} $v_{\sigma}$ to lie within the axion window.

\ack This work is partially supported by the {\it Cluster of Excellence  PRISMA}  DGF no. EXC 1098 and the DFG research grant HO 4166/2-1.
W.S.~is supported by the ERC Advanced Grant SPLE under contract ERC-2012-ADG-20120216-320421, by the grant FPA2012-32828 from the MINECO, and the grant SEV-2012-0249 of the ``Centro de Excelencia Severo Ochoa" Programme.

\vspace{4mm}


\end{document}